# Evaluation of insulating behavior by Pb Substitution and Pressure-Induced Superconductivity in $La_2O_2Bi_{3-x}Pb_{1+x}S_6$

Rajveer Jha[1,*], Valenta Jaroslav[1], Naohito Tsujii[1], Takashi Naka[2], Takeshi Kawahata[3], Chiduru Kawashima[3], Hiroki Takahashi[3], Takao Mori[1,4] and Yoshikazu Mizuguchi[5,*]

[1]WPI Center for Materials Nanoarchitectonics (WPI-MANA), National Institute for Materials Science (NIMS), Namiki 1-1, Tsukuba 305-0044, Japan.
[2]National Institute for Materials Science, Research Center for Functional Materials, 1-2-1, Sengen, Tsukuba, Ibaraki 305-0047, Japan.
[3]Department of Physics, College of Humanities and Sciences, Nihon University, Setagaya, Tokyo 156-8550, Japan.
[4]Graduate School of Pure and Applied Sciences, University of Tsukuba, Tennodai 1-1-1, Tsukuba 305-8671, Japan.
[5]Department of Physics, Tokyo Metropolitan University, 1-1 Minami-Osawa, Hachioji, Tokyo 192-0397, Japan.

E-mail: JHA.rajveer@nims.go.jp, mizugu@tmu.ac.jp

## Abstract

We report synthesis, characterization, and physical properties of layered bismuth-sulfide compounds $La_2O_2Bi_{3-x}Pb_{1+x}S_6$. We synthesized a new $La_2O_2Bi_3PbS_6$ compound, whose crystal structure is similar to those of the $La_2O_2Bi_3AgS_6$ superconductor or $La_2O_2Bi_2Pb_2S_6$ (insulator) with a four-layer-type conducting layer. The crystal structure of $La_2O_2Bi_3PbS_6$ is characterized in a tetragonal $P4/nmm$ space group, and the lattice parameters are $a$ = 4.084(1) Å and $c$ = 19.71(2) Å. The obtained $La_2O_2Bi_3PbS_6$ sample exhibits narrow gap semiconductor (metallic-like) transport behavior with an anomaly near the 160 K. A metallic-like to insulator transition has been observed through Pb substitution, with increasing $x$ in $La_2O_2Bi_{3-x}Pb_{1+x}S_6$. Furthermore, we applied external pressure for $x$ = 0 and observed pressure-induced superconductivity. The onset of superconductivity appeared at 0.93 GPa, and the highest transition temperature was 3.1 K at 2.21 GPa.

## 1. Introduction

In order to achieve superconductivity with exotic pairing mechanisms and/or a high transition temperature ($T_c$), layered compounds are promising candidates: for example, high-$T_c$ cuprates, Fe-based superconductors, $BiS_2$-based superconductors, and so on [1-6]. The typical $BiS_2$-based superconductors, $REO_{1-x}F_xBiS_2$ ($RE$ represent the rare-earth element) and $Sr_{1-x}La_xFBiS_2$ with $T_c$ of 2–11 K, have two $BiS_2$ conducting layers in between their blocking layers [4-19]; here, we call those $BiS_2$-based superconductors two-layer-type. Recently, we have reported superconductivity in a related layered compound $La_2O_2Bi_3AgS_6$ with a superconducting transition temperature ($T_c$) of ~0.5 K [20]. The crystal structure of $La_2O_2Bi_3AgS_6$ is composed of $La_2O_2$ blocking layers similar to two-layer-type compounds, but their conducting layer is four-layer-type $M_4S_6$ ($M$ = Ag, Pb, Bi) layers; hence, we call those compounds four-layer-type.

The highest record of $T_c$ in the four-layer-type La$_2$O$_2$$M_4$S$_6$ compounds is 4.0 K at ambient pressure, which has been achieved by tuning chemical pressure through substitutions at the La, $M$, and/or S sites [21-23].

The first report of four-layer-type compound was on La$_2$O$_2$Bi$_2$Pb$_2$S$_6$ (reported as LaOPbBiS$_3$) with a narrow gap semiconductor and insulating transport behavior was observed at low temperatures [24]. Initially, the crystal structure of La$_2$O$_2$Bi$_2$Pb$_2$S$_6$ was reported as the stacks of the $M_4$S$_6$ conducting layers with random occupation of Bi and Pb at the $M$ site. [24]. After the first synthesis report, the structural analyses revealed that the $M_4$S$_6$ conducting layer is the stacks of two BiS$_2$ layers and rock-salt-type Pb$_2$S$_2$ (connected two PbS sheets) layer [25]. The crystal structure of La$_2$O$_2$$M_4$S$_6$ can be achieved by insertion of rock-salt-type M-S layer into the van der Waals gap between two BiS$_2$ layers of REOBiS$_2$ [26]. Along with the strategy, La$_2$O$_2$Bi$_3$AgS$_6$ was synthesized by inserting AgBiS$_2$ layer into LaOBiS$_2$ and superconductivity was observed [20]. Recently, the effects of external pressure on superconducting properties have been studied for the four-layer-type La$_2$O$_2$Bi$_3$Ag$_{0.6}$Sn$_{0.4}$S$_6$ [27], La$_2$O$_2$Bi$_3$AgS$_6$ [28,29] and La$_2$O$_2$Bi$_2$Pb$_2$S$_6$ [29] compounds, and the highest $T_c$ of ~8 K was achieved for La$_2$O$_2$Bi$_3$Ag$_{0.6}$Sn$_{0.4}$S$_6$ at 6.4 GPa [27]. Under pressure, $T_c$ was enhanced up to 4.1 K at 3.1 GPa for the La$_2$O$_2$Bi$_3$AgS$_6$, and the possible charge density wave (CDW) like hump at $T^* \sim$ 155 K was simultaneously suppressed with pressure [29]. In addition, pressure-induced superconductivity was observed in La$_2$O$_2$Bi$_2$Pb$_2$S$_6$, which was achieved by the suppression of local disorder [29].

In this study, we have synthesized the La$_2$O$_2$Bi$_3$PbS$_6$ compound ($M$ = PbBi in a formula of La$_2$O$_2$Bi$_2$$M_2$S$_6$). In addition, we have synthesized La$_2$O$_2$Bi$_{3-x}$Pb$_{1+x}$S$_6$ ($x$ = 0.1, 0.5, 1.0) samples and observed that the unit cell noticeably expanded with increasing Pb concentration $x$ in the rock-salt-type $M$-S layers. The temperature dependence of electrical resistivity [$\rho$(T)] shows that the La$_2$O$_2$Bi$_3$PbS$_6$, and La$_2$O$_2$Bi$_{2.9}$Pb$_{0.1}$S$_6$ ($x$ = 0, 0.1) compounds are narrow gap semiconductor (metallic-like) with anomalies around $T^* \sim$ 160 K. In contrast, the samples with $x$ = 0.5, 1.0 exhibits semiconducting (insulating) behavior at low temperatures. The change can be regarded as a metallic-like to insulator transition by Pb substitution in La$_2$O$_2$Bi$_{3-x}$Pb$_{1+x}$S$_6$. Although we did not observe superconductivity in La$_2$O$_2$Bi$_3$PbS$_6$ down to $T$ = 1.8 K at ambient pressure, superconductivity was observed in La$_2$O$_2$Bi$_3$PbS$_6$ under high pressure. The highest $T_c$ in La$_2$O$_2$Bi$_3$PbS$_6$ was $T_c$ = 3.1 K at 2.21 GPa.

## 2. Experimental details

The polycrystalline samples of La$_2$O$_2$Bi$_{3-x}$Pb$_{1+x}$S$_6$ ($x$ = 0, 0.1, 0.5, 1.0) were prepared by a solid-state reaction method. The powders of La$_2$S$_3$ (99.9%), Bi$_2$O$_3$ (99.9%), Pb (99.99%), and grains of Bi (99.999%) and S (99.99%) with a nominal composition of La$_2$O$_2$Bi$_{3-x}$Pb$_{1+x}$S$_6$ were mixed in a pestle and mortar, pelletized, sealed in an evacuated quartz tube, and heated at 720 °C for 15 h. To obtain homogeneous samples, the sintered samples were reground, pelletized, and heated at 720 °C for 15 h. The phase purity of the prepared samples and the optimal annealing conditions were examined by X-ray diffraction (XRD) by using RINT TTR-3 diffractometer (Rigaku Co., Akishima, Tokyo, Japan) employing Cu-K$_\alpha$ radiation. The crystal structure parameters were refined using the Rietveld method with RIETAN-FP [30]. The schematic image of the crystal structure was drawn using VESTA [31]. The actual compositions of the synthesized samples were investigated by using energy-dispersive X-ray spectroscopy (EDX) with TM-3030 (Hitachi). The electrical resistivity down to $T$ = 1.8 K was measured by the four-probe technique on the Physical Property Measurement System (PPMS: Quantum Design). Pressure experiments were performed using a pressure cell HPC-33 produced by Electro Lab Corporation. The pressure cell was installed into the PPMS for standard four-probe measurement of resistance. Daphne oil 7373 was used as a pressure transmitting medium, and pressure was determined by a superconducting transition of Pb manometer (purity: 4N) [32]. The high-pressure XRD experiments were performed at room temperature using Mo-K$_\alpha$ radiation on a Rigaku (MicroMax-007HF) rotating anode generator equipped with a 100 µm collimator. The diamond anvil cell was used for the high pressure XRD experiment. Daphne 7474 was used as a pressure-transmitting medium.

## 3. Results and discussion

Figure 1 shows the Rietveld-refinement profile obtained from powder XRD pattern of La$_2$O$_2$Bi$_3$PbS$_6$ at room temperature. The XRD data is well fitted with a $P4/nmm$ model with a resulting reliability factor of $R_{wp}$ =7.81%. As shown in the inset of Fig. 1, the unit cell of La$_2$O$_2$Bi$_3$PbS$_6$ is composed of stacked $M_4$S$_6$ layers and fluorite-type La$_2$O$_2$ layers, and the $M_4$S$_6$ layer is composed of the BiS$_2$-type layers and (Bi,Pb)S layers. The Rietveld analysis results suggest that La$_2$O$_2$Bi$_3$PbS$_6$ has smaller lattice parameters than those of La$_2$O$_2$Bi$_2$Pb$_2$S$_6$ [24,25].

We have examined the effects of Pb substitution for the Bi site in La$_2$O$_2$Bi$_{3-x}$Pb$_{1+x}$S$_6$ ($x$ = 0, 0.1, 0.5, 1.0) samples. Figure 2(a) shows the powder XRD pattern of La$_2$O$_2$Bi$_{3-x}$Pb$_{1+x}$S$_6$ at room temperature. All the obtained samples are crystallized in the tetragonal structure with the space group of $P4/nmm$. The lattice parameters increase with increasing $x$ in La$_2$O$_2$Bi$_{3-x}$Pb$_{1+x}$S$_6$. As the Pb concentration increases from $x$ = 0 to 1.0, the 006 peak shifts towards the lower 2$\theta$ angle side, which suggests systematic lattice expansion by Pb substitution. Also, 110 peak broadening suggests the existence of in plane local distortion due to Pb substitution in the $La_2O_2Bi_{3-x}Pb_{1+x}S_6$. We observed small impurity peaks of $Bi_2S_3$ for $x = 0, 0.1$ and $La_2O_2S$ impurity peak for $x = 1.0$. The $La_2O_2S$ impurity peak is indicated by asterisks (*) in Fig.2(a). The shift of the 006 and 110 peak position is shown in Fig. 2(b, c). The parameters obtained by the Rietveld refinements for $La_2O_2Bi_{3-x}Pb_{1+x}S_6$ are summarized in Table 1.

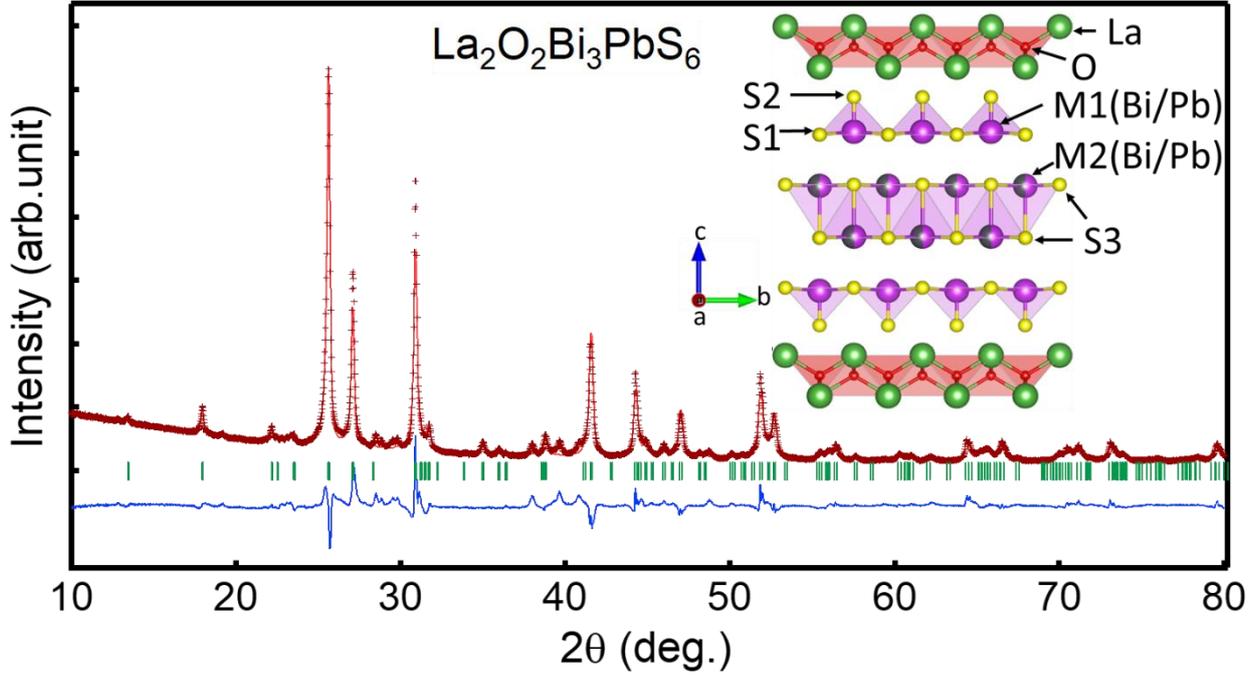

**Figure 1**: Powder XRD pattern and the Rietveld fitting for the $La_2O_2Bi_3PbS_6$ sample. The inset is the schematic image of crystal structure of $La_2O_2Bi_3PbS_6$ drawn with parameters obtained from the Rietveld analysis.

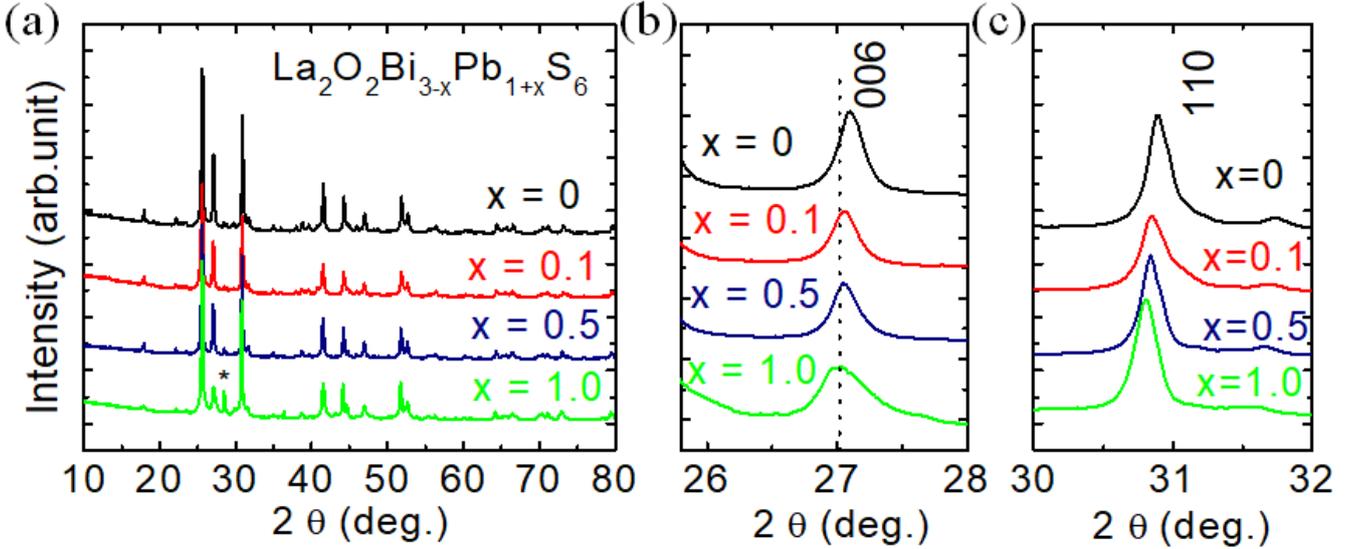

**Figure 2**: (color online) (a) Powder XRD patterns for all the samples of $La_2O_2Bi_{3-x}Pb_{1+x}S_6$. The peak of the $La_2O_2S$ impurity phase was indicated by an asterisk. (b) Zoomed profiles near the 006 peaks. (c) Zoomed profiles near the 110 peaks. The dashed line is an eyeguide.

Table 1: Rietveld-refined lattice parameters for $La_2O_2Bi_{3-x}Pb_{1+x}S_6$

| Samples | $x = 0$ | $x = 0.1$ | $x = 0.5$ | $x = 1.0$ |
|---|---|---|---|---|
| $a$ (Å) | 4.084(1) | 4.093(2) | 4.097(1) | 4.097(2) |
| $c$ (Å) | 19.71(2) | 19.74(1) | 19.76(3) | 19.79(1) |
| $R_{wp}$ % | 7.81 | 8.29 | 8.77 | 13.11 |

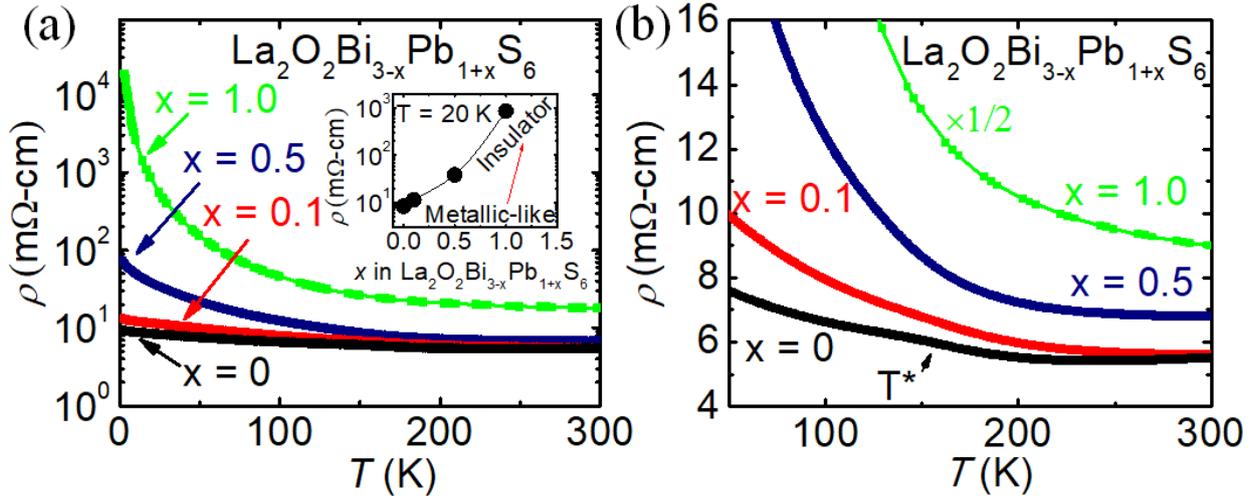

**Figure 3**: (color online) (a) Temperature dependences of electrical resistivity at 300-1.8 K for $La_2O_2Bi_{3-x}Pb_{1+x}S_6$ ($x$ = 0, 0.1 0.5, 1.0), inset is the doping dependence of $\rho$ at T = 20 K. (b) $\rho(T)$ curve in the temperature range of 300-60 K.

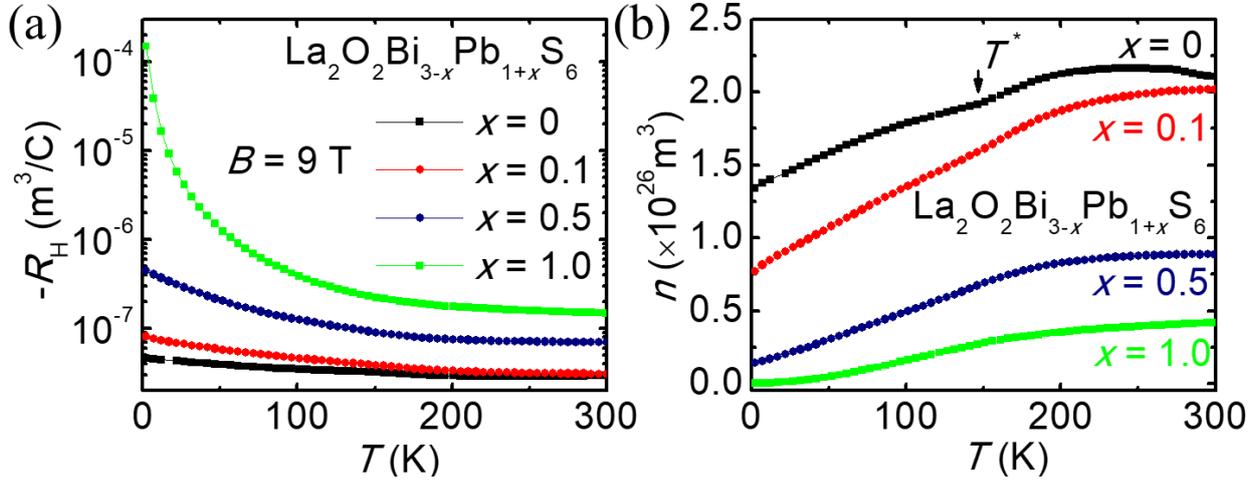

**Figure 4**: (a) Temperature dependences of Hall coefficient ($R_H$) for $La_2O_2Bi_{3-x}Pb_{1+x}S_6$ ($x$ = 0, 0.1 0.5 and 1.0) under the magnetic field 9 T. (b) Temperature dependences of estimated carrier concentration for $La_2O_2Bi_{3-x}Pb_{1+x}S_6$ ($x$ = 0, 0.1 0.5 and 1.0).

Figure 3(a) shows the temperature dependence of electrical resistivity for the $La_2O_2Bi_{3-x}Pb_{1+x}S_6$ samples. $La_2O_2Bi_3PbS_6$ ($x$ = 0) shows the lowest electrical resistivity among all the obtained samples. We observed a CDW-like hump near the anomaly temperature ($T^*$) of ~160 K in the $\rho(T)$ curve for $x$ = 0, as shown in Fig. 3(b). The hump observed in the $\rho(T)$ curve is quite similar to that observed for the $La_2O_2Bi_3AgS_6$ and $EuFBiS_2$ superconductors [20,33]. The anomaly at $T^*$ could be associated with the localization of electrons due to disorder or strong electron-electron interactions near $T$ ~160 K. The anomaly at $T^*$ exists up to $x$ = 0.1, but it is clearly smaller than that for $x$ = 0. Semiconducting behavior in the $\rho(T)$ curve was observed for $x$ = 0.5 and 1, and the anomaly at $T^*$ is completely suppressed for $x$ = 0.5 and 1. The $\rho$ for $x$ = 0.5 and 1 remarkably increases with decreasing temperature, which is consistent with a previous report on semiconducting (insulating) behavior at low temperatures [24]. The rapid increase in $\rho$ can be understood with a decrease in in-plane chemical pressure by Pb substitution from analogy to two-layer-type systems [34]. Another factor is the loss of electron doping due to the Pb substitution at Bi site considered as a narrow gap semiconductor-insulator transition. The trend can be regarded as a metallic-like to insulator transition by substitution when plotting the doping dependence of $\rho$ [inset of Fig. 3(a)], which is similar to metal-insulator transitions typically observed in $d$-electron oxides systems [35-40]. The structural distortion as a significant factor in metal-insulator transition has also been reported in various studies [40,41]. In the current $BiS_2$-based systems, insufficient in-plane chemical pressure results in local structural distortion [42,43], and the local distortion would be originating from the presence of Bi (and Pb as well, in this system) 6s lone pairs [44]. Therefore, the metal-insulator transition induced by Pb substitution in $La_2O_2Bi_{3-x}Pb_{1+x}S_6$ would provide a new platform to study metal-insulator transition phenomena in *p*-electron systems.

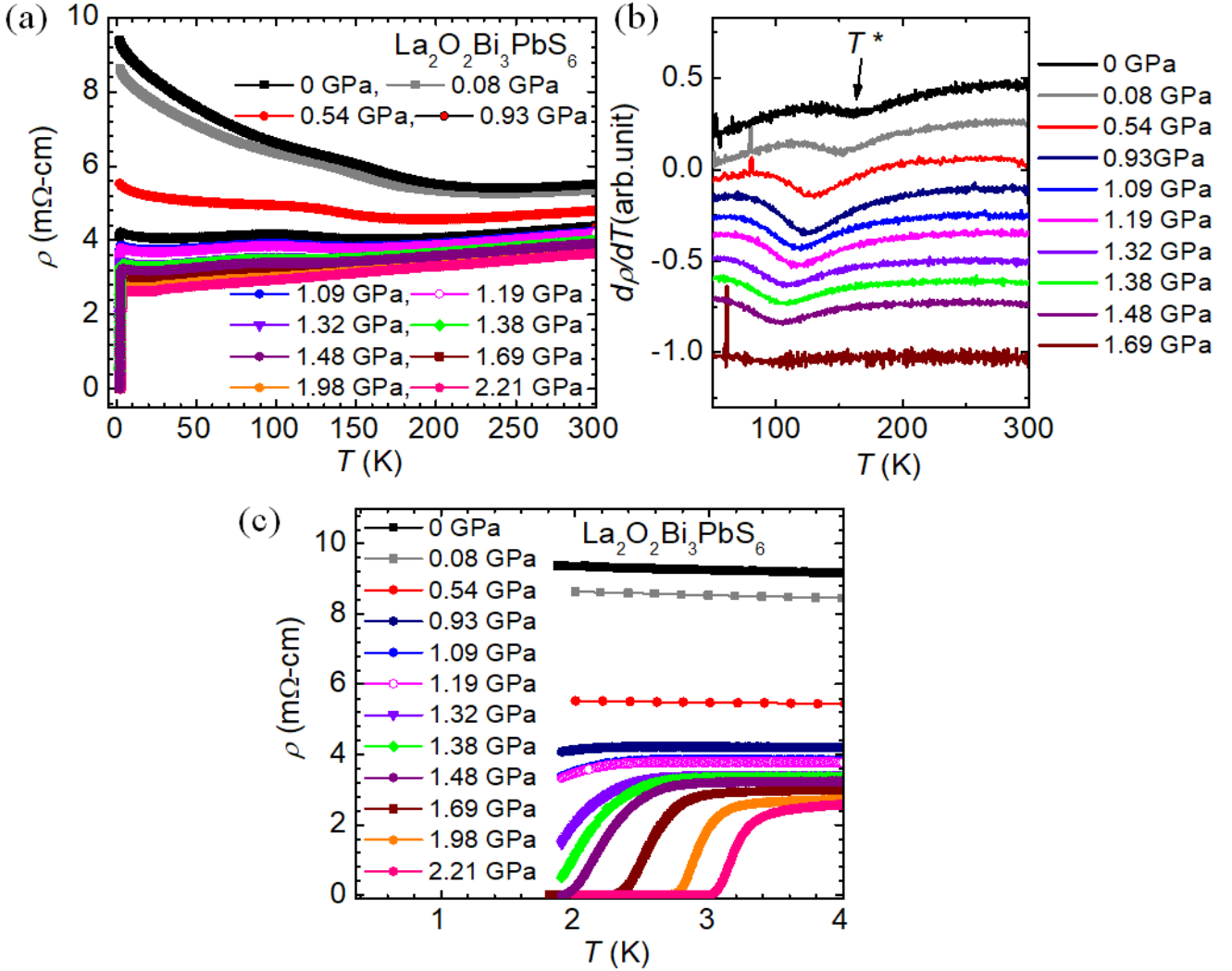

**Figure 5**: (color online) (a) Temperature dependences of electrical resistivity from 300-1.8 K for $La_2O_2Bi_3PbS_6$ at various pressures. (b) The first derivative of resistivity with respect to temperature of sample for selected pressure points. The color-coding of (b) is the same as in (a); vertical arrow mark the $T^*$ (c) $\rho(T)$ curve in the temperature range of 4 -1.8 K.

To examine the change in carrier concentration, the Hall coefficient was measured for $La_2O_2Bi_{3-x}Pb_{1+x}S_6$ ($x$ = 0, 0.1, 0.5, 1.0). Figure 4(a) shows the temperature dependence of Hall coefficient $-R_H(T)$ for $La_2O_2Bi_{3-x}Pb_{1+x}S_6$. The measurements were performed under magnetic fields ($B$) of 9 T at T = 2–300 K. We observed negative Hall coefficients for all the samples, which suggests the electron-type carriers dominate in the whole temperature range. The absolute value of $R_H$ increases with increasing Pb concentration in $La_2O_2Bi_{3-x}Pb_{1+x}S_6$, which indicates a decrease in electron carrier concentration by Pb substitution. We calculated carrier concentration ($n$) by assuming a single-band model $n$ = $1/(eR_H)$; similar analyses were used for other Bi-layered compounds [24, 45-47]. Figure 4(b) shows the temperature dependence of $n$ for $La_2O_2Bi_{3-x}Pb_{1+x}S_6$. We observed an anomaly near ~160 K for $x$ = 0, which is consistent with resistivity anomaly. The $n$ for $x$ = 0 is 2.22×10$^{26}$ m$^{-3}$ for the $La_2O_2Bi_3PbS_6$ and decreases with increasing Pb concentration. The large carrier concentration for

$La_2O_2Bi_3PbS_6$ compared to that of $La_2O_2Bi_2Pb_2S_6$ suggests that electron carriers are created by Bi-rich composition because of the difference of valence of $Bi^{3+}$ and $Pb^{2+}$. The substitution of $Pb^{2+}$ for $Bi^{3+}$ in $La_2O_2Bi_{3-x}Pb_{1+x}S_6$ can have significant effects on its electronic properties, particularly in terms of electron doping and the resulting impact on the observed insulating behavior. Likewise, the substitution of $Pb^{2+}$ for $Bi^{3+}$ in a material can lead to a loss of electron doping and an increase in insulating behavior due to changes in electronic structure, band structure, chemical bonding, disorder, and ion size. These factors collectively reduce the availability and mobility of charge carriers, limiting the material's conductivity. In addition, the Pb substitution expands the *ab* plane, which results in a decrease in in-plane chemical pressure [34]. Then, local structural disorders [43] become remarkable and carriers are localized by Pb substitution.

On the basis of the transport results at ambient pressure, we decided to examine the effects of high pressure on transport properties to check pressure induced superconductivity in $La_2O_2Bi_3PbS_6$. Figure 5(a) shows the temperature dependence of electrical resistivity for $x = 0$ at different pressures from 1.8 to 300 K. Electrical resistivity decreases with pressure. The semiconducting-like behavior observed at ambient pressure gradually suppressed with pressure, and metallic behavior was observed at $P = 1.98$ and 2.21 GPa. Semiconducting to metallic transition by pressurizing has been reported for a two-layer-type $BiS_2$-based $Sr_{0.5}La_{0.5}FBiS_2$ superconductor [15]. Noticeably, the first-principles calculation suggested that the charge density wave instability in two-layer-type $LaO_{0.5}F_{0.5}BiS_2$ [48]. We consider that the appearance of the metallic conductivity under pressure is linked to the possible modification of the CDW-like states with $T^*$ even for four-layer-type compound. Also, the anomaly at $T^*$ in $\rho(T)$ curve at ambient pressure for $x = 0$ shifts towards the lower temperatures with pressure. Though, the link between the anomaly at $T^*$ and the superconductivity remains unclear due to the absence of a microscopic evidence for the CDW. Figure 5(b) demonstrates the evolution of $T^*$ under pressure, which was estimated from the minimum of $d\rho/dT$ curve, as indicated by vertical arrow. The anomaly in $\rho(T)$ is completely suppressed at high-pressure range of $P \geq 1.98$ GPa. In Fig. 5(c), the low temperature $\rho(T)$ near the superconducting transition is plotted. The $T_c^{onset} = 1.9$ K appears at $P = 0.93$ GPa. The zero-resistivity temperature of $T_c^{zero} = 1.9$ K was observed at $P = 1.48$ GPa. The $T_c$ increases with pressure and reaches a maximum of 3.1 K at $P = 2.21$ GPa. The $T_c$ continuously increases with pressure up to our measurement limit.

Figure 6 shows the temperature-pressure phase diagram for $x = 0$. The obtained results of pressure dependencies of $T_c$ and $T^*$ for $x = 0$ are summarized in the phase diagram, which clearly indicates the correlation between $T_c$ and $T^*$. As the pressure increases, the $T^*$ decreases monotonically with increasing pressure up to 1.69 GPa. Superconductivity with $T_c^{zero} = 1.75$ K and anomaly ($T^*$) coexist at $P = 1.69$ GPa. With further pressuring, $T^*$ disappears, and $T_c$ increases with increasing pressure. The maximum $T_c^{zero}$ of 3.1 K was observed at $P = 2.21$ GPa. We could not observe a sign of saturation in $T_c$ up to the pressure of 2.21 GPa due to the pressure limit of our system. The trend of pressure dependences of $T_c$ and $T^*$ for $x = 0$ in the present study are similar to those observed in the $La_2O_2Bi_3AgS_6$ compound.

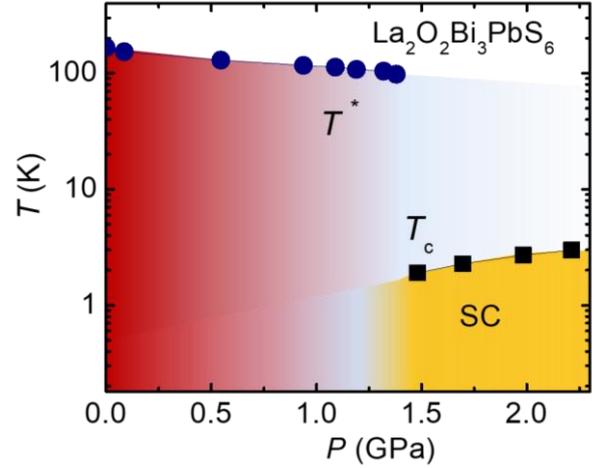

**Figure 6**: (color online) (a) Temperature–pressure phase diagram for $La_2O_2Bi_3PbS_6$. The values of $T^*$ and $T_c$ are extracted from the resistivity data shown in figure 5, and SC symbolizes superconductivity.

The temperature-pressure phase diagram reported by Aulestia *et al.* for Ag-rich $La_2O_2Bi_3AgS_6$ [28] exhibits qualitative similarities to Fig. 6. As a fact, $La_2O_2Bi_3PbS_6$ and $La_2O_2Bi_3AgS_6$ exhibit common features: the existence of pressure-induced superconductivity with the suppression of the semiconductor-like behavior, $T^*$ is suppressed with pressure, and $T_c$ increases as $T^*$ decreased by applying pressure [29]. Obtained results for the pressure dependencies of $T_c$ and $T^*$ for $x = 0$ are consistent with the results reported for pressure studies on $La_2O_2Bi_3AgS_6$ and $La_2O_2Bi_2Pb_2S_6$ by Y. Yuan et al. [29].

Figures 7(a) show the temperature dependence of electrical resistivity under various magnetic fields [$\rho(T,B)$] for $x = 0$ at $P = 2.21$ GPa. With an increasing magnetic field, $T_c$ shifts towards a lower temperature side. With the gradual shifts in $T_c$ with increasing applied magnetic field, the fundamental phenomena of superconductivity have been confirmed for $x = 0$ at $P = 2.21$ GPa, with which we confirmed that the observed resistivity drops under pressure were originated from the superconductivity. For the applied magnetic field of 5 T, the superconducting states are completely suppressed. The magnetic field-temperature phase diagram is plotted in Fig. 7(b). The upper critical fields [$B_{c2}(T)$] are estimated based on the temperature where resistivity becomes 50% of the normal-state resistivity near $T_c$, and $\rho_n$ = 50% is marked as $T_c^{mid}$ in Fig. 7(a). The $B_{c2}(T = 0)$ was analyzed using the conventional one-band Werthamer-Helfand-Hohenberg (WHH) model [49], which gives $B_{c2}(0) = -0.693T_c(dB_{c2}/dT)_{T=T_c}$. $B_{c2}(0)$ is 1.15 T for $x$ = 0 at $P$ = 2.21 GPa.

To investigate the evolution of crystal structure under high pressure for $La_2O_2Bi_3PbS_6$, we performed high-pressure XRD at room temperature. Figure 8(a) shows the XRD patterns taken at high pressures, and the zoomed profiles near the (116) and (200) peaks are displayed in Fig. 8(b). As revealed by the Rietveld refinement shown in Fig. 1, the ambient-pressure crystal structure was refined by the tetragonal (P4/nmm) model. The tetragonal phase is maintained up to 2.72 GPa. The increase in $T_c$ observed in the high-pressure resistivity measurements is therefore understood by the effect of lattice shrinkage in the tetragonal system. Similar increase in $T_c$ by pressure in the tetragonal phase was observed in $La_2O_2Bi_3Ag_{0.6}Sn_{0.4}S_6$ [27], and the origin of the increase in $T_c$ was explained by the in-plane chemical pressure effect [34, 50]. The shift of the (200) peak to a higher angle by pressure, zoomed in Fig. 8(b), suggests the in-plane shrinkage. Hence, we consider that the same scenario can be applied to understand the increase in $T_c$ in $La_2O_2Bi_3PbS_6$ examined in this study.

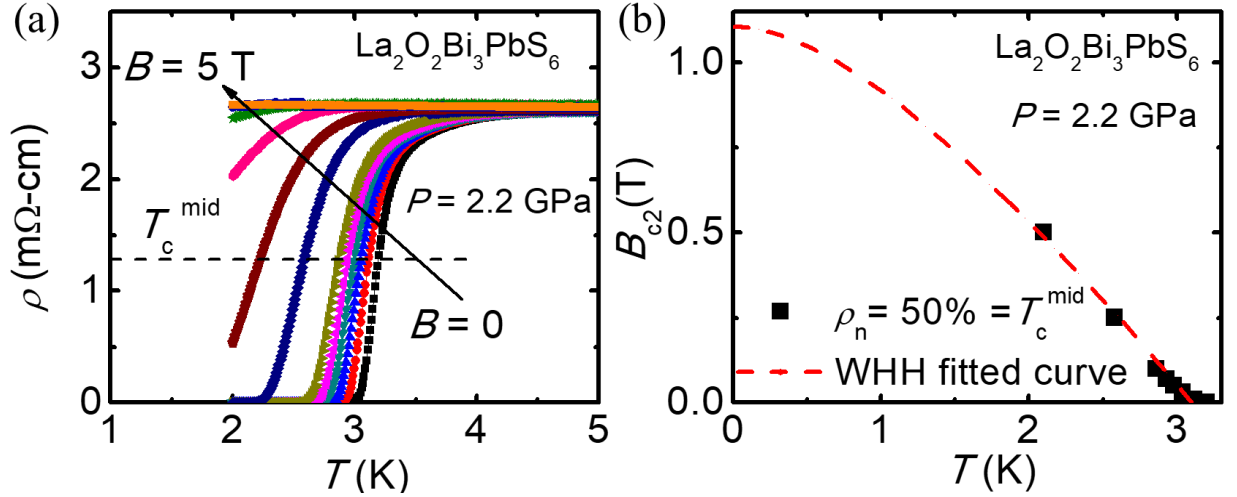

**Figure 7**: (color online) (a) Temperature dependences of electrical resistivity under magnetic fields from 5.0-1.9 K for $La_2O_2Bi_3PbS_6$ at 2.21 GPa pressure. (b) Temperature dependences of $B_{c2}(T)$ for $La_2O_2Bi_3PbS_6$. Experimental data points are shown with symbols, and fitting curves by the WHH theory are shown with dotted lines.

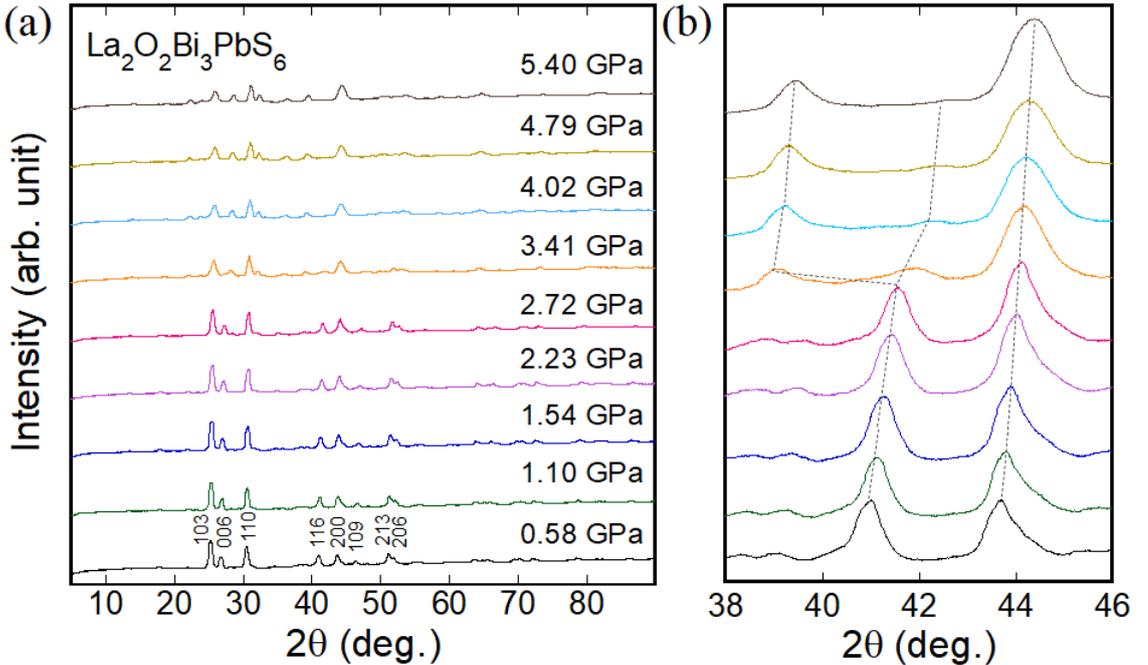

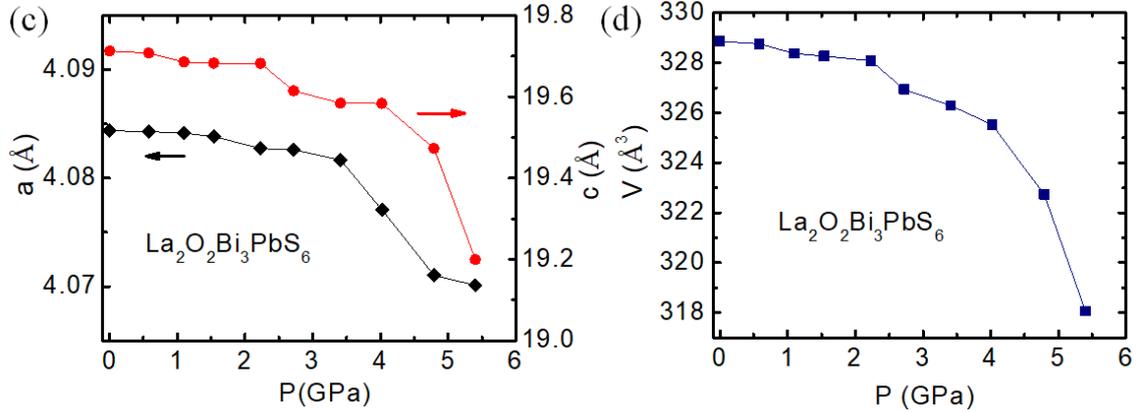

**Figure 8**: (color online) (a) XRD patterns collected at various pressures for $La_2O_2Bi_3PbS_6$. (b) The zoomed profiles near the (116) and (200) peaks. (c) The evolution of the lattice parameters a, and c as a function of pressure (d) unit-cell volume as a function of pressure.

At pressures above 3.41 GPa, the (116) peak splits into two peaks as shown in Fig. 8(b), and the (200) peak exhibits broadening. These changes in the XRD pattern above 3.41 GPa indicate lowering symmetry from tetragonal. In addition, we noticed that the (006) peak shows a clear shrinkage between 2.72 and 3.41 GPa. This suggests that the structural transition would associate with a large shrinkage along the $c$ axis. We have refined XRD pattern at each pressure and the refined lattice parameters a, c and unit cell volume as a function of pressure shown in Fig. 8(c,d). The Volume of the unit cell is decreasing rapidly above $P = 3.41$ GPa suggests structural transition occurred above this pressure. Although the structural analysis by synchrotron XRD for the higher-pressure phase is needed, we expect a further increase in $T_c$ in the present material under higher pressures because $T_c$ of $BiS_2$-based superconductors commonly shows a jump-like increase by structural symmetry lowering [51].

## 4. Summary

We have obtained a new Bi-rich $La_2O_2Bi_3PbS_6$ compound, crystallizing in the four-layer-type $La_2O_2M_4S_6$ structure. The $x = 0$ sample exhibits the metallic-like behavior with an anomaly at $T^* = 160$ K in $\rho(T)$. With increasing Pb concentration in $La_2O_2Bi_{3-x}Pb_{1+x}S_6$, the $\rho(T)$ gradually turns into a semiconducting-like behavior. Thus, Pb substitution in $La_2O_2Bi_{3-x}Pb_{1+x}S_6$ induces a metallic-like to insulator transition. The shift from metal to localized states are explained by the changes in carrier concentration and local structural disorder. We studied the external pressure effects on $x = 0$ and observed pressure-induced superconductivity in $x = 0$. The anomaly temperature $T^*$ in $\rho(T)$ decreased with pressure, and the $T_c$ appears around $P = 0.93$ GPa. The highest $T_c$ ($\rho = 0$) is 3.0 K at $P = 2.21$ GPa. The high-pressure XRD suggests that the increase in $T_c$ observed in this study is explained by the shrinkage of the $ab$-plane, and further increase in $T_c$ is expected at a higher pressure in the distorted crystal structure.

**Acknowledgements**

This work was partially supported by JST Mirai Program Grant JPMJMI19A1. Author (R.J.) is grateful to the support by the Grand-in-Aid from Japanese Society for the Promotion of Science (JSPS), KAKENHI, No. 20K22331. Author (V.J.) is thankful to the support by Grand-in-Aid from JSPS, KAKENHI, No. 21F21322.